# International environmental treaties: An honest or a misguided effort

Reza Hafezi [a,b], David A. Wood [c], Firouzeh Rosa Taghikhah [d,b]

[a] Science & Technology Futures Studies, National Research Institute for Science Policy (NRISP), Tehran, Iran.

[b] Sydney Environment Institute, The University of Sydney, Australia.

[c] DWA Energy Limited, Lincoln, United Kingdom.

[d] Business School, University of Sydney, Australia.

**Abstract:**

Climate change and environmental concerns represent a global crisis accompanied by significant economic challenges. Regular international conferences held to address these issues, such as in the UK (2021) and Egypt (2022), spark debate about the effectiveness and practicality of international commitments. This study examines international treaties from a different perspective, emphasizing the need to understand the power dynamics and stakeholder interests that delay logical actions to mitigate anthropogenic contributions to climate change and their impacts. Environmental and social concerns tend to increase within nations as their economies develop, where they fight to keep acceptable standards of living while reducing emissions volume. So, nations play disproportionate roles in global decision-making based on the size of their economies. Addressing climate change requires a paradigm shift to emphasize acknowledging and adhering to global commitments through civil pressure, rather than relying on traditional yet biased systems of international political diplomacy. Here, climate-friendly actions are evaluated and ideas to promote such activities are proposed. We introduce a "transition regime" as a solution to this metastasis challenge which gradually infects all nations.

**Keywords:** Climate change; international commitments; sustainable development; transition regime; more feasible routes to net zero.

1. Introduction:
1.1. More than Just a Political Game?

Climate change and environmental concerns pose significant challenges to sustainable development at both global and national levels [1]. As human activities continue to impact the planet's ecosystems and climate, there is an urgent need for collective action to mitigate these effects. The perspective presented seeks to address a central research question: Are international environmental agreements effective in achieving their goals, or do economic interests ultimately undermine their success? Some answers and solutions are proposed to tackle the global challenges related to climate change, which endangers the survival of mankind and a large portion of the life that currently exists on Earth in the long run. This work offers perspective addressing a real conundrum, analogous to that of the frog placed in gradually heated water (referring to the tale of the boiling frog). It is crucial to make decisions and take actions, otherwise, we will rightly be held accountable by future generations (i.e. history) for the damage knowingly caused to our planet. Sitting on the fence or procrastinating cannot be justified any longer, difficult decisions leading to positive climate impacts need to be taken, and in fact are now long overdue. We cannot steal other generations' futures and leave a permanently damaged environment as our legacies.

International environmental agreements have evolved over time, with their roots primarily in the 1970's and 1980s [2], when certain countries began to recognize the need for collective action to address environmental challenges. Since then, numerous agreements have been adopted to address a wide range of issues, from ozone depletion to biodiversity loss. The driving forces behind these agreements include increasing awareness of environmental issues, scientific advancements leading to better understanding of the complicated network of factors that influence regional and global climates, and the recognition that individual countries or sectors of society cannot effectively tackle these problems on their own.

Sustainable development is defined as meeting the needs of the present without compromising the ability of future generations to meet their own needs. Industrialization, on the other hand, refers to the process of transforming an economy from an agricultural base to one based on manufacturing, service provision and other industries. Technological advancements refer to the development and adoption of new technologies, which can have both positive and negative consequences for the environment and climate.

The industrial revolution, human desire for increased prosperity, less arduous life styles, modern industrial and technological concepts, and the pursuit of global economic domination have resulted in negative consequences for the environment and many of its ecosystems [3]. Over the decades, some radical scientists and activists have emphasized the destructive environmental consequences of many industrial applications of technological advancements. For example, Theodore John Kaczynski (alias Unabomber), a convicted urban terrorist, took extreme direct and unjustified violent actions against industrialization (Washington Post, 1995). On the other hand, numerous legitimate attempts have been made to alert and unite nations in addressing environmental issues synergistically, committing the international community to take tangible steps to mitigate the negative consequences of industrialization and excessive consumerism. These efforts have led to fragmented groups of nations working together, as evidenced by the multinational attendance and pledges made at events such as the Rio Earth Summit in 1992, Kyoto Protocol in 1997, Rio+20 in 2012, the Paris Agreement (COP21), Glasgow Climate Change Conference (COP26; November 2021), the Sharm el-Sheikh Climate Change Conference (COP27; November 2022) and other similar accords.

However, it is impossible to claim that such accords have been successful, as negative climate and ecosystem damage related to industrial growth and growing fossil-fuel energy consumption related to population growth continues unabated. A cynical view suggests that these accords have merely resulted in signatures from senior politicians representing many nations, while their respective countries continue with business-as-usual strategies driven by national economic interests, causing tangible negative environmental consequences. To date, these accords have failed to deliver substantial improvements to the environment or noticeable slow down climate-change impacts. This is exemplified by the positive impacts on air quality and carbon dioxide ($CO_2$) emissions reduction observed in 2020 due to COVID lockdowns and forced curtailments of industrial activities in numerous countries, as well as a similar outcome in 2009 following the global banking crisis-induced recession.

Another drawback of multinational accords is their fragility and lack of long-term binding commitments, allowing countries to withdraw from them (e.g., the United States' exit from the Paris Agreement) or dilute their commitments based on national interests promoting economic growth. Economic growth typically necessitates increased energy consumption, and as the energy mix for the most populous nations is still dominated by fossil fuels, this inevitably leads to greater atmospheric pollution [4]. Consequently, the altruistic objectives of preserving favourable conditions on Earth for ecosystems to thrive are undermined by national economic growth drivers. At the level of international environmental agreements, it is clear that "money talks," especially for some nations, and, unfortunately, at the expense of others. Thus, a dichotomy exists between environmental aspirations and economic growth objectives in many countries. This study will consider potential counterarguments and alternative perspectives on the effectiveness of international environmental agreements, with the aim of providing a more balanced and realistic analysis.

### 1.2. Conflicts of Interests: The Source of Challenges

A particularly challenging issue lies in how developed, developing, and underdeveloped nations communicate on matters of global concern. When negotiating multinational agreements, each nation typically seeks to promote its own interests and maximize the economic benefits it might accrue. Game theory suggests that sustainable multinational policy agreements are more likely to be reached if win-win conditions for all parties can be achieved. However, such theoretical "games" often assume that all participants have equal power, and/or fail to take account of differing long-term and short-term priorities. In multinational climate response agreements, it is apparent that some participants exert minimal influence, especially those nations contributing minimally to the global economy. Conflict arises when the world's powers gather to limit their own development, resulting in what could be described as a theatre of climate change mitigation.

It is instructive to examine the sources of historical cumulative $CO_2$ emissions from countries responsible for the largest emissions reported from 1850 to 2021 (Figure 1). Two main contributions can be identified: (1) $CO_2$ emissions related to fossil fuel consumption (grey bars); and (2) $CO_2$ emissions related to land-use change and deforestation (green bars). From 1950, fossil fuel-related $CO_2$ emissions increased about 3.85 times, from 6.5 Gt in 1950 to 25.1 Gt in 2000. In contrast, land-use-change-related $CO_2$ emissions experienced about a 22% decrease from about 5.5 Gt in 1950 to 4.3 Gt in 2000. These statistics underscore the significance of fossil fuels and increased energy demand for economic and population growth in their contributions to $CO_2$ emissions growth.

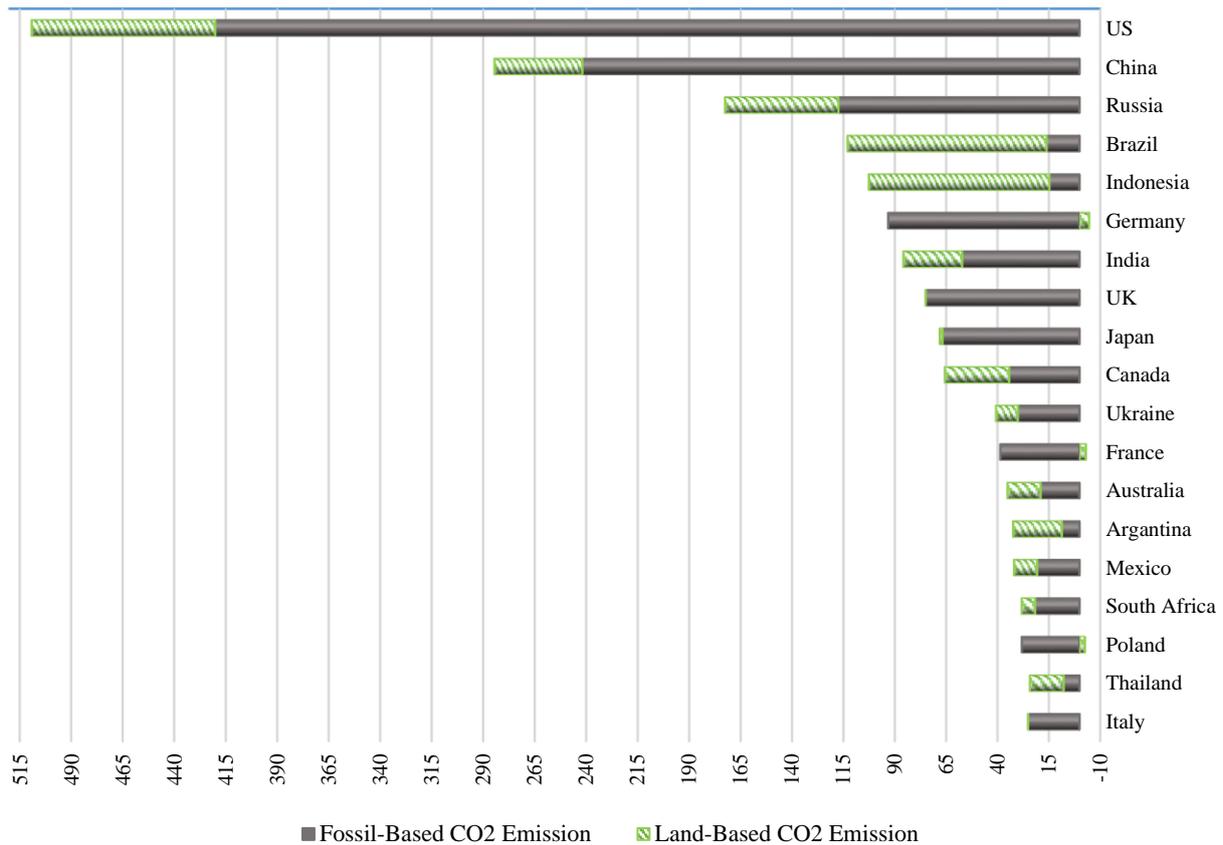

**Figure 1. Cumulative $CO_2$ emission from 1850 to 2021 for countries with the largest cumulative emissions (billions of tonnes). Note: when a green bar appears right side of the grey bar it means that the corresponding nation experienced negative land-based $CO_2$ emission.**

National contributions to $CO_2$ emissions are not as clear-cut as the Figure 1 statistics suggest. A more relevant approach involves analyzing national contributions to emissions in a more holistic manner. One such method is to incorporate a life-cycle analysis related to the production and consumption of key commodities and services. This would involve considering the extraction of raw materials at one end of the life cycle along various supply chains, through processing, manufacturing, and distribution, to post-consumption recycling of waste materials (termination of life cycle). Thus, emissions recorded by resource-rich nations (through extraction and distribution processes) should not be attributed solely to the producing nations but also partially shared with those nations consuming, transporting and trading the commodities and services. Recording net or life-cycle contributions to $CO_2$ emissions would provide a more accurate recognition of the nations most responsible for such emissions.

We need to consider a less discussed issue, that is related to the question: What share of carbon emissions is allocated to early-stage technology development? Historically, what we called the "Developed World" and leading companies (through radical innovation processes) spent vast resources and contributed over long periods of time to environmental pollution. However, technology and knowledge development drive national/regional economic and political power, it also increases the level of public access to products and services without reinventing the wheel and emitting carbon. Fairly, such emissions (caused by human activities that are not constrained to confined regions/nations/races etc.) should not be assigned the responsibility of just a few industrial-pioneering countries. On the other hand, it cannot simply be divided based on a simple per capita allocation to calculate nations' shares. Since many nations are still fighting for access to many resources and industrial technologies, it is unfair for them to be financially burdened with the historical pollution associated with resources and technologies they have never had access to.

It is important to address potential counterarguments or alternative perspectives regarding the conflicts of interest in international environmental agreements. Critics may argue that focusing solely on economic solutions, such as $CO_2$ emission taxes, may not be sufficient to address the complex and multifaceted nature of climate change. Additionally, some may contend that implementing an international fiscal system may be challenging due to issues of sovereignty and enforcement, and the reality that some nations will game such a system. By acknowledging these perspectives, the analysis becomes more balanced and comprehensive.

Another factor concerns how the economically developed world should support developing and underdeveloped nations to help them achieve targeted economic growth and improve their living standards. This includes facilitating access to sufficient, relatively low-cost, clean energy sources. In the current world, many nations, particularly developing ones, opt for the cheapest or most accessible energy solutions. Unfortunately, many of these solutions are the least environmentally friendly (e.g. coal without carbon capture). The position taken, and the modifications demanded to the final COP26 statement by India exemplify how conflicts of interest can lead to lobbying, attenuation, refusal, and even withdrawal from international environmental commitments.

Such conflicts of interest primarily arise from the economic drivers influencing national policy decisions. This suggests that economic solutions, such as $CO_2$ emission taxes, are required. The fairest way to implement such a fiscal solution would be to use net (life-cycle) emissions, with the proceeds of the taxes raised used to provide financial and technical assistance to less developed nations, thereby helping them meet environmental commitments while combating poverty by ensuring a certain level of economic growth. To implement such a fiscal system, fair and unbiased international regulators with enforcement powers would be necessary. Moreover, this system would need to be accountable and demonstrate tangible emissions reduction benefits. Such systems are, of course, of no use if they just work on paper and/or are subject to manipulation by vested interests.

### 2. Political Representation of Civil Society

In an ideal world, governing politicians should represent the collective desires of the population in promoting environmental, cultural, and economic aspirations. However, political oligarchies (represented by political parties in many countries) often attempt to preserve the interests of their supporters and financial backers. Democratically elected politicians tend to prioritize short-term issues, such as job creation or economic growth, as their policies and decisions related to such issues can achieve tangible short-term rewards. For example, the United States' withdrawal from the Paris Agreement in 2017 prioritized U.S. economic interests over long-term environmental concerns [5]. This example highlights that national environmental policies are often driven and/or constrained by economic influencers and vested interests to secure their political survival.

Another factor impacting international environmental accords, such as the Paris Agreement or the Kyoto Protocol, is that not all sectors of the global population have the same level of influence in terms of the scope they cover. The agendas of such accords typically determine the issues that are prioritized, which often do not incorporate all the relevant issues affecting various sectors of society. For instance, the Kyoto Protocol focused on reducing greenhouse gas emissions but did not adequately address deforestation or the needs of indigenous communities. To address these gaps in representation, it is essential to involve marginalized groups, or those with limited economic influence, and ensure their voices are heard in the decision-making process.

On the other hand, many environmentally friendly rules and regulations around the world, such as the Clean Air Act in the United States or the European Union's Emissions Trading System, have been prompted by activists highlighting the detrimental environmental effects of unbridled industrial development justified solely on economic grounds. This suggests that in order to promote more rapid and effective actions to limit future environmental damage, a greater influence should be assigned more broadly to civil society and the verified data analysis of scientists working out with the funding and influence of governments and industry. This could act as a check on the economically driven decisions promoted by political leaders and parties, provided it can be achieved without causing civil unrest, anarchy, and negative impacts on economic development. The success of the "Fridays for Future" movement in mobilizing millions of young people worldwide to demand climate action exemplifies the power of activism [6]. However, activists often face challenges in the form of opposition from powerful industries, established political hierarchies, and scepticism from the public. Environmental activism in many nations is currently linked to organizations expressing extreme political views, which makes the public suspicious of their long-term political objectives. Nevertheless, bottom-up actions and demands, given an appropriate voice and influence, can help balance top-down decisions myopically clouded by a focus unjustifiably skewed towards economic vested interests.

To achieve a paradigm shift that promotes bottom-up, climate-change mitigation activism and representation while avoiding political extremism, widespread public support around the world is needed. This should focus on balancing the requirements for prompt environmentally friendly actions with investments targeting more sustainable economic development and growth. Specific mechanisms or strategies that could be employed to achieve this include policy changes, such as implementing carbon pricing; institutional reforms that broaden the scope and strengthen the powers of environmental agencies; and the formation and support of civil society organizations dedicated to environmental sustainability. While these strategies have the potential to bring about positive change, they may also face resistance from entrenched interests and require careful negotiation to avoid unintended consequences.

Unfortunately, there is temporal heterogeneity in political decision-making. The consequences of political decisions tend to emerge on different time scales. For example, economic consequences often materialize in the short or medium term following a political decision. On the other hand, the consequences of environmental and social-welfare-directed political decisions typically emerge in the long run. The delayed response to the ozone depletion crisis in the 1980s illustrates the negative impact of short-term thinking on environmental efforts [7]. To foster a more long-term perspective among decision-makers, it is crucial to prioritize education and awareness about the importance of sustainability and environmental stewardship.

The prevailing commercial systems depend on consumerism to maintain economic growth at a stable pace. Almost everyone is part of the commercial consumption process. It is not possible to change behaviours easily because our interests are, in the short term, served by those consumptions. Consequently, we are all contributing at some level to preserving the business-as-usual status quo. The fashion industry's impact on water pollution and natural resource depletion exemplifies the link between consumerism and environmental problems [8]. Potential strategies for encouraging more sustainable consumption habits include promoting circular economy principles and raising consumer awareness about the environmental impacts of their purchasing choices.

From another perspective, the global environmental crisis presents an opportunity to refocus humanity on a sustainable future and reconnect individuals with their environments. Crises can and often do, ultimately ignite revolutions in attitudes, life styles and individual/group aspirations as well as political objectives [9]. The global environmental crisis is a compelling reason to shift the current paradigm that is endangering life on Earth. Like political revolutions, civil society needs to replace the dominant and myopic economic-political preferences of most nations with politics striving for sustainable growth. Such a shift should help promote a high quality of human life and preserve ecosystems for future generations. However, achieving such a shift will not be easy due to the existing vested interests that represent formidable obstacles.

To overcome these challenges, individuals, governments, and organizations must collaborate and pursue innovative solutions that promote sustainable consumption and production patterns. Public education and awareness campaigns can play a critical role in fostering a culture of sustainability and environmental responsibility. Additionally, governments can create incentives for businesses to adopt greener practices and invest in clean technologies.

### 3. Transition: Reality or Illusion
### 3.1. (Un)Sustainability and Political Interests

Political systems and their relationship with sustainability have faced criticism due to the complex nature of balancing various interests and the long-term consequences of policy decisions. Major challenges related to sustainability and political systems include addressing the negative impacts of climate change, fostering sustainable development, and managing the trade-offs between short-term gains and long-term environmental and social goals. In his book titled "Sustainability," Portney investigates the concept of sustainability, focusing on urban sustainability programs, and analyzes a range of case studies through qualitative and quantitative methodologies [10].

To tackle unsustainability social movement is crucial which is represented in a political context as democracy. Interestingly, Chapman et al. showed that civil interpretation of social complexity and understanding of democracy complexity significantly affect how they demand and support democracy [11]. So experiencing democracy in the real world increases the possibility of demanding democracy and governmental responsibility (e.g. transparency) [12]. Although regime type influences democracy and demand for it at the individual level, it has been proved that individual activities and support for democracy are important predictors of the democratization of society regardless of regime type [11].

Cities represent the largest part of society where climate change clearly displays its negative consequences. For instance, climate change-induced extreme weather events and sea-level rise is likely to result in mass migrations from many low-lying coastal cities (although in most cases such Hurricane Katrina migrations were just temporary, still uncertainties are existing). Additionally, increased healthcare costs can be attributed to urban air pollution, which exacerbates respiratory and cardiovascular diseases [13]. Portney [11] highlights concerns about sustainability motivating actions based on limiting or reducing consumption, defining consumption and waste reduction as the ultimate solutions to improve sustainability. However, resistance has emerged in response to these measures.

The first source of resistance is initiated and supported by developing and less developed economies that are on the path to improving their quality of life. Global treaties, such as the Paris Agreement, force these nations to slow down development, at least in its current form, which are likely to create some tensions between the goals of economic growth and environmental sustainability.

The second source of resistance is rooted in the economy. Sustainable development requires the full participation of economic parties, as economic directions affect human well-being in the long run. From a politician's perspective, there is a trade-off between choosing to follow optimal future-oriented decisions (e.g., $CO_2$ taxes) and satisfying short-term social and economic needs, which are more likely to result in future re-election. This dilemma is exemplified by the U.S. withdrawal from the Paris Agreement in 2017, prioritizing economic interests over global climate goals [5].

While the solution may appear simple, the path toward sustainability is difficult due to the need to consider multiple agents with various interests and even significant conflicts. To address these challenges, policymakers must engage in open dialogue with stakeholders, develop flexible strategies to balance economic growth and environmental concerns and transparently prioritize long-term sustainability in decision-making processes.

### 3.2. Need for Transition Regime: Giants as Game Changers

After the cold war, it was assumed that the world would develop toward liberalisation and democratic hegemony, while global events of the past two decades revealed that optimism was premature. Indeed, many nations experienced an erosion of democracy and an increase in authoritarianism during that period [14]. Moreover, there is a heterogeneity in nations' power where some countries can make international obligations. For instance, referring to Putnam's two-level game [15, 16], U.S. negotiation behaviours and other international outcomes can be interpreted and conceptualised based on U.S. dynamics of preferences over time [17]. On the other hand, when there is no collective willingness/desirability to cooperate globally, as an example to mitigate climate change, an international (i.e. climate) regime will be developed only through a small set of influencial actors, acting as a hegemony [18-20]. One key actor is the U.S., exerting influence both directly and indirectly through its international alliances [17]. Another key actor is China, also exerting influence both directly and indirectly through its international alliances. The currently polarized political and economic ideologies of these two key actors, supported by their own various alliances make the problem of developing an effective global climate-change-mitigation accord even more difficult and complex to achieve. Layered on top of this political/economic complexity, is the diverse nature of climate change environmental and non-environmental impacts being faced by specific nations around the world, some with higher temperature increases, droughts, wildfires etc. (e.g. greater than the global average) and some with economic consequences eroding their standards of living.

In such circumstances, an effective transition regime can only be achieved if a number of dominant actors join to develop and support a climate regime. Snidal proposed the K-group theory where K demonstrates the number of dominant nations that support the regime [21]. The optimum number for such group theory is one (e.g. K=1), however as long as one nation can manage collective actions regardless of the states in the group. The K-group theory discusses why U.S. and China have become less willing to cooperate since 2015 [17]. It seems likely that annual global greenhouse gas emissions have not yet peaked. For example, global atmospheric $CO_2$ concentrations and sea level continue to rise [22, 23]. As time goes in without those emissions being abated the greater the environmental consequences will be for the planet. Hence, the need to initiate a transition regime to adjust to climate change requires urgent consideration.

### 4. Conclusion: Ideas for Addressing the Global Environmental Crisis

Evidence shows that climate change is not a conspiracy theory. Environmental obligations determined by multinational conventions are important but insufficient to mitigate the environmental impacts of climate change in a timely manner. Lifestyles and human activities across the world remain driven by self-interest and tend to be dominated by the relentless pressure to achieve short-term economic growth through unsustainable consumerism dependent on unrealistic levels of energy consumption. As Antonio Guterres (Secretary-General of the United Nations) said, "It's time to go into emergency mode," particularly when humanity and the biosphere as a whole are in critical danger. The question remains: what measures should be taken as part of that emergency mode, and how should decisions be made? The following factors should be considered:

- *Do not let inertia or vested interests be the enemy of progress.* Tangible transitions to sustainable lifestyles must start immediately, even at a slow pace. This initiative must begin at the individual level, as there is no time to wait for governments to act, given that many have inappropriate vested interests creating barriers to action.
- *Investigate potential win-win solutions that benefit the population in general.* Modern energy sources and $CO_2$ management systems are controversial issues for resource-rich economies and nations striving to move their populations out of poverty. More balanced measures of contributions to pollution should be based on a life-cycle analysis of commodity production and consumption, allowing the costs of mitigating emissions to be shared more proportionately between commodity-supplier nations and commodity-consuming nations.
- *Bottom-up pressure is more effective.* It is crucial to manage a global paradigm shift to sustainable lifestyles and promote, from the ground level, commitments to international environmental obligations, regardless of short-term political considerations.

- *We are systematically destroying large sectors of the biosphere, not the Earth itself.* It is essential to inform society that our current actions have severe negative consequences for the human race and many ecosystems, not the planet. The Earth will survive with or without us and many other life forms. Slogans create shared visions, and shared visions shape our mindset. The human race is following a path that can be realistically described as long-term suicide. About 50 years ago, scientists and social activists were worried about the next generation; now, many sectors of society are dealing with the consequences of unsustainable development as part of daily life.
- *Implement awareness-raising with a comprehensive and even-handed evaluation system (framing normative messages).* Statistics can be manipulated to some extent to portray politically acceptable trends. What metrics are measured, where they are measured, how transparently they are reported and how they are interpreted matter. Governments tend to define indexes to support their vested interests. An unbiased and transparent protocol is essential for defining and assessing indexes and determining how nations contribute to or capture pollution. This requires an impartial regulator and independent international observer with sufficient media access to publish findings that will, from time to time, make unpleasant reading for certain governments.
- *Reinvent the wheel.* It is apparent that the concepts of sustainability and climate-change mitigation evolve over time. All previous efforts have strived unsuccessfully to adapt to climate-change impacts. Efforts have been made to align with negative consequences and to manage current and future risks (e.g., be reactive instead of proactive). However, globally we now need to mitigate what is not plausible through existing international procedures and accords. The new world's challenges require innovative solutions and managing paradigms. Old-fashioned routes and failed initiatives will not lead us to new destinations.
- *Develop innovation platforms to facilitate the initialization of ecosystem preservation.* Climate change and natural disasters are large-scale problems that require outside-the-box solutions (e.g., changes in patterns/trends and thinking/cultural paradigms) instead of inside-the-box solutions (e.g., projects, new tools, and fixing existing problems). Both product and system innovations offer the potential to significantly contribute to the Earth's future. Now, governments and international agencies should develop platforms to facilitate and regulate agents' roles using games and initiatives rather than expressing disfunctional representative roles, which we are experiencing today. It is apparent that collectively the world's national governments are currently misrepresenting the voices of their respective societies.
- *Develop the transition regime.* Nations play different roles and consequently, they are not commited to shared rules. The transition regime is proposed as a facilitator to dampen environmental threats. Regarding K-group theory, it is crucial to develop a new K-group with U.S. and China cooperating at its core. To be effective that new K-group would need to promote transparency and achieve the collective support for a more rigorous climate-change-mitigation policy through its various alliances.

Ideally, the human race needs to transition from the "as is" lifestyle and redefine the quality of life in more sustainable terms, where habitability and the lives of all species and future generations are given consideration and respect. The voices of individuals and civic institutions against narrow-minded and culpable economic interests must be strengthened. The footprint of industrial-economic development of nations and their contributions to the environmental crisis must be laid bare for all to see and respond to. Moreover, the developed world should play its part in empowering less developed countries to create a more balanced and sustainable global political-economic system. To tackle environmental concerns, we need to avoid tunnel vision focused primarily on carbon emissions and the economic development constraints inhibiting meaningful responses. It is necessary to concentrate more on stimulating short-term sustainable human lifestyles, matched with appropriate longer-term economic development aspirations. If we fail collectively to do this, the opportunity to prevent a long-term nightmare scenario for the biosphere from unfolding will be lost.

It is not the planet that needs saving, but a significant portion of the life that exists upon it. As human beings, we must collectively establish new visions to protect the lives of future generations of all species, instead of adopting reductionist views that abstractly and less effectively attempt to save the Earth. The Earth does not need our sympathy.